\documentclass[%
 twocolumn,
superscriptaddress,
 amsmath,amssymb,
 prl,
]{revtex4-1}

\usepackage{graphicx}% Include figure files
\usepackage{dcolumn}% Align table columns on decimal point
\usepackage{bm}% bold math
\usepackage{xcolor}
\newcommand{\rt}[1]{\textcolor{black}{#1}}                            % rt: red text     % rt: red text
\usepackage{soul}
\usepackage[normalem]{ulem}
\newcommand{\rsmath}[1]{\bgroup\markoverwith{\textcolor{red}{\rule[0.5ex]{2pt}{0.4pt}}}\ULon {\textcolor{red}{#1}}}                                           %rsmath: red stroke for math environment

\usepackage{eso-pic}
\AddToShipoutPictureBG*{%
\AtPageUpperLeft{%
\hspace{\paperwidth}%
\raisebox{-\baselineskip}{%
}}}%

\begin{document}

\title{Designing superselectivity in linker-mediated multivalent nanoparticle adsorption}

\author{Xiuyang Xia}
\thanks{\rt{Present address: Arnold Sommerfeld Center for Theoretical Physics and Center for NanoScience, Department of Physics, 
    Ludwig-Maximilians-Universit\"at M\"unchen, Theresienstraße 37, D-80333 M\"unchen, Germany}}
\affiliation{School of Chemistry, Chemical Engineering and Biotechnology, Nanyang Technological University, \\62 Nanyang Drive, 637459, Singapore}

\author{Ran Ni}\email{r.ni@ntu.edu.sg}
 \affiliation{School of Chemistry, Chemical Engineering and Biotechnology, Nanyang Technological University, \\62 Nanyang Drive, 637459, Singapore}%Lines break automatically or can be forced with \\


\begin{abstract}
Using a statistical mechanical model and numerical simulations, we provide the design principle for the bridging strength ($\xi$) and linker density ($\rho$) dependent superselectivity in linker-mediated multivalent nanoparticle adsorption. When the bridges are insufficient, the formation of multiple bridges leads to both $\xi$- and $\rho$-dependent superselectivity. Whereas, when the bridges are excessive, the system becomes insensitive to bridging strength due to entropy-induced self-saturation and shows a superselective desorption with respect to the linker density. Counterintuitively, lower linker density or stronger bridging strength enhances the superselectivity.
These findings enhance understanding of relevant biological processes and open up opportunities for applications in biosensing, drug delivery, and programmable self-assembly.
%These findings not only help in understanding relevant biological processes but also open up opportunities for applications in biosensing, drug delivery, and programmable self-assembly.
\end{abstract}

\maketitle
  
Using multiple weak bindings to form effectively strong interactions, namely multivalent interactions, constitutes the cornerstone of various biological processes~\cite{mammen1998polyvalent,kitov2003nature,huskens2006multivalent}. %TODO: more
 They also play a crucial role in determining the fate of nanomedicines in biological media: multivalent nanoparticles target cell surfaces overexpressing receptors, e.g., cancer cells or viral infections, while leaving others untouched~\cite{davis2010evidence,koenig2021structure}. 
This ``on-off'' binding can be conceptualized as \emph{superselectivity}, which refers to the targeted adsorption probability $\theta$ responding superlinearly to the receptor density $n_r$: $ {\rm d} \log  \theta / {\rm d} \log {n_r}  > 1 $~\cite{martinez2011designing}.
 
Compared to conventional direct bindings, indirect linker-mediated bridging provides new axes for regulating interactions by adjusting both the density and specificity of linkers. 
Here \emph{linkers} refer to small molecules with two ends that can transiently and reversibly bind with receptors and ligands coated on hosts and guest objects, respectively, forming bridges to link them. 
Many biological processes involve this interaction, e.g., protein degradation and transport~\cite{schapira2019targeted}, biomolecular condensates formation~\cite{rosenzweig2017eukaryotic,rayman2018tia,bouchard2018cancer,ryu2021bridging} and chromatin fiber reshaping~\cite{malhotra2021unfolding}, etc. 
Moreover, the linker specificity offers the \emph{in situ} programmability that specifically ``sews''  a pair of building blocks by adding corresponding linkers in artificial multicomponent systems, e.g., DNA origami~\cite{rothemund2006folding}, DNA linked colloids~\cite{biancaniello2005colloidal,oleg1,oleg2,oleg3,girard2019particle,lowensohn2019linker,xia2020linker,Tavares2020}, artificial protein-interaction networks~\cite{sanders2020competing,heidenreich2020designer,nandi2022affinity} and vitrimers~\cite{lei2020entropy,xia2022entropy}.

With these new design elements, linker-mediated interactions also exhibit the superselectivity originating from multivalency. 
This scheme serves as the foundation for modern sandwich immunoassays and DNA array detections~\cite{voller1976enzyme,taton2000scanometric,cao2002nanoparticles,ja021096v,lukatsky2004phase}.
Moreover, linker-mediated systems undergo an entropy-driven re-entrant transition with increasing linker density~\cite{lowensohn2019linker,xia2020linker}. Higher linker density promotes the bridge formation, but excessive density ``over-charges'' the system, preventing the formation of bridges. 
This process, recently found to be superselective, involves the membrane repair protein annexin A5 binding to anionic lipid membranes assisted by the linker Ca$^{2+}$\cite{curk2022controlling}.
Nevertheless, due to the complex phase behavior observed in linker-mediated systems, a universal design principle for inducing superselectivity remains elusive.
Here combining the mean-field theory of linker-mediated interactions~\cite{xia2020linker} with the Langmuir-like multivalent adsorption model~\cite{martinez2011designing,angioletti2017exploiting}, we study the bridging strength and linker density dependent superselectivity in the linker-mediated multivalent nanoparticle adsorption, offering a comprehensive guideline for designing the superselectivity in linker-mediated systems.
\begin{figure}[!t]
\centering
        \includegraphics[width=0.48\textwidth]{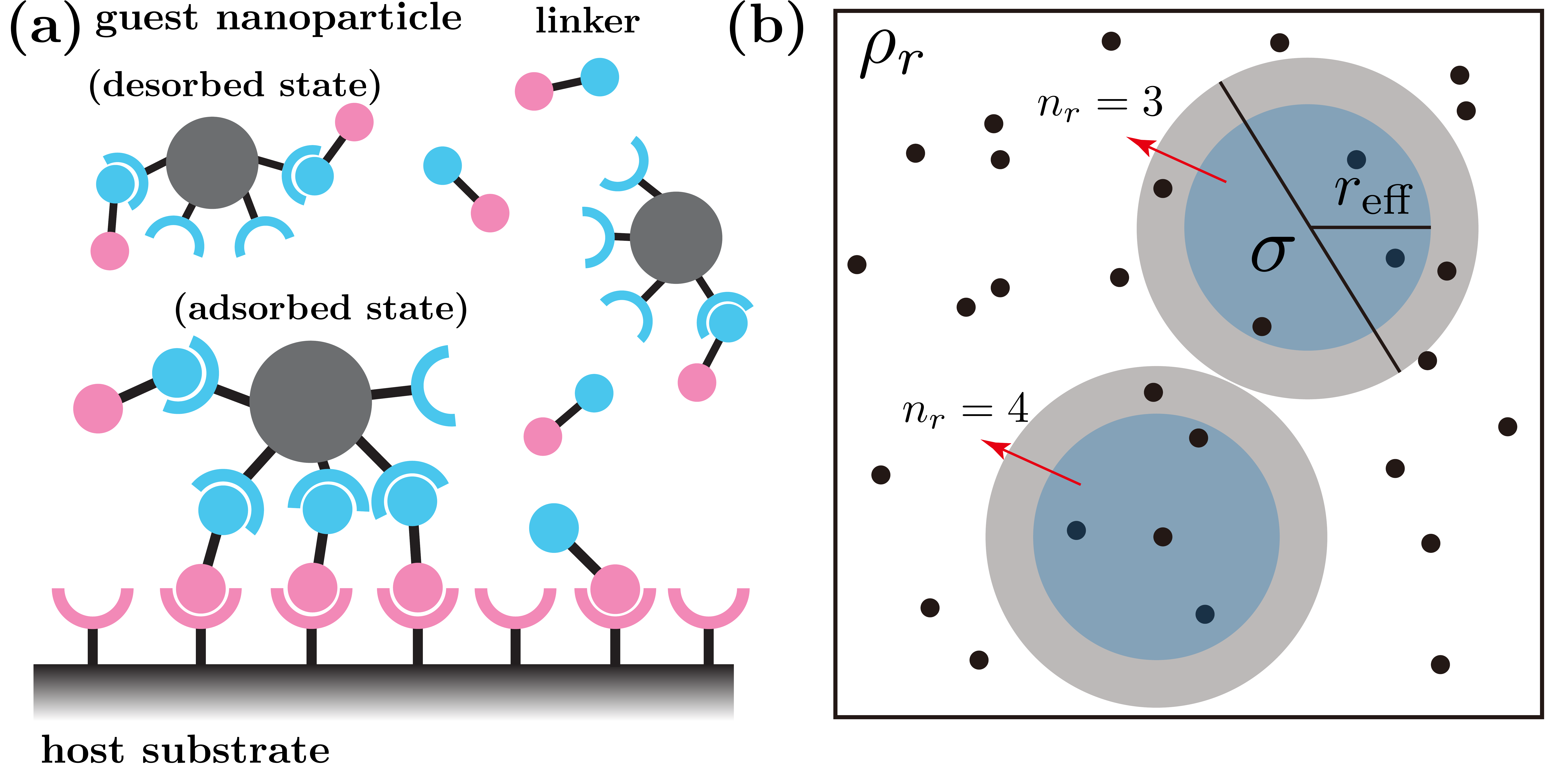}
    \caption{\label{fig1} \textbf{Linker-mediated adsorption of multivalent nanoparticles.} (a) Schematic representation of multivalent nanoparticles adsorption composed of a guest-linker-host sandwich. (b) Illustration of the GCMC simulation.	 The black dots are immobile receptors with grey circles being adsorbed nanoparticles. Receptors within blue regions can form bridges with the ligands on nanoparticles mediated by implicit linkers.}
\end{figure}

\paragraph*{Model.}
We consider a prototypical coarse-grained model involving guest nanoparticles being adsorbed on a host substrate (Fig.~\ref{fig1}a). 
Rather than direct bindings~\cite{martinez2011designing}, indirect bridgings mediated by bivalent free linkers in solution occur between the ligands on nanoparticles and the receptors on host substrate.
We define a nanoparticle \emph{adsorbed} when it forms at least one bridge with the host substrate, and  \emph{desorbed} when no bridge is formed and it remains in the bulk solution. 
Each nanoparticle is coated with $n_l$ mobile ligands, i.e., the ligands can move on the particle surface.
Linkers are modeled as an ideal gas of density $\rho = e^{\beta\mu}/\Lambda^3$ with chemical potential $\mu$, the de Broglie wavelength $\Lambda$, and \(\beta=1 / k_{B} T\), where $k_B$ and $T$ are the Boltzmann constant and temperature of the system, respectively.
We consider that a site with \(n_r\) receptors on the host substrate can adsorb at most one guest particle. The mobility of ligands allows each ligand to potentially bridge with every receptor on the site, ensuring the maximum combinatorial entropy and the validity of the mean-field approximation~\cite{angioletti2014mobile}.

We assume that the two ends of linkers are specifically bound with receptors and ligands, and the partition functions $\xi_r= e^{-\beta f_r}$ and $\xi_l= e^{-\beta f_l}$ with the binding free energy $f_r$ and $ f_l$, respectively. The formation of a bridge causes a mean-field conformational entropy cost $\Delta S_{\rm cnf}$ with the partition function $\xi_{\rm cnf}=e^{k_B^{-1}\Delta S_{\rm cnf}}$.
Taking the desorbed state as the reference state, the adsorbed state partition function is defined as $q= e^{-\beta F_{\rm eff} }-1$, where the effective interaction $F_{\rm eff}$ is the free energy difference between the system with adsorbed and desorbed guests~\cite{angioletti2017exploiting}. 
With the saddle-point approximation, we obtain \rt{(see SM Sec.~I~\cite{supinfo2})}
	\begin{equation}\label{eq:F_eff}
	\begin{aligned}
	        \beta F_{\rm {eff}} =& n_l \log \left(\frac{{n}_l-b}{{n}_l}\right) +  n_r \log \left(\frac{{n}_r-b}{{n}_r}\right) +b,
	\end{aligned}
    \end{equation}
with the equilibrium number of bridging linkers
\begin{equation}\label{eq:b}
	b = \frac{\left(n_l+n_r+\Gamma^{-1} \right) - \sqrt{\left(n_l+n_r+\Gamma^{-1} \right)^2-4n_ln_r}}{2}.
\end{equation}
Here, we define the relative bridging strength 
\begin{equation}\label{eq:gamma}
    \Gamma = \frac{\xi_l\xi_r\xi_{\rm cnf}\rho}{1+(\xi_l+\xi_r)\rho+ \xi_l\xi_r\rho^2},
\end{equation}
which represents the bridging strength with respect to all unbridged states: both empty, only one side occupied by linkers and both occupied by linkers, similar to the ligand-receptor ``affinity'' in Ref.~\cite{curk2022controlling}.

We assume that the adsorption on each substrate site is independent with no interaction between particles adsorbed on adjacent sites. The grand-partition function averaging over all sites is $\Xi = \langle 1+z_g q \rangle _{\langle n_r \rangle}$, where \(z_g\) represents the guest nanoparticle activity in the bulk solution, proportional to the particle density and adsorption volume~\cite{martinez2011designing}. \( \langle \cdot \rangle_{\langle n_r \rangle} \) calculates the average over the Poisson distribution of receptors with the mathematical estimate $\langle n_r \rangle$.
Thus, the probability of the nanoparticle adsorption follows the Langmuir adsorption isotherm~\cite{martinez2011designing,langmuir1918adsorption}
\begin{equation}\label{eq:theta}
	\theta :=\frac{\partial \log \Xi} { \partial \log z_g} = \left\langle \frac{{z_gq}}{{1+z_gq}} \right\rangle_{\langle n_r \rangle}.
\end{equation}
	
We perform grand-canonical Monte Carlo (GCMC) simulations with implicit ligands and linkers to verify Eq.~\ref{eq:theta} \rt{(see SM Sec.~IV~\cite{supinfo2} and \cite{xia2023role,wang2012selectivity,phan2023bimodal,dubacheva2015designing,frenkel2023understanding})}. 
As shown in Fig.~\ref{fig1}b, we model the multivalent nanoparticles as hard disks of diameter $\sigma=1$ and volume $s_{\rm hd}=\pi/4$ controlled by the chemical potential $\mu_{g}$. We consider the explicit receptors are immobile under a Poisson distribution with number density $\rho_r$ on a 2D substrate. 
 A receptor can only bridge with mobile ligands on the nanoparticles within a center-to-particle-center distance $r_{\rm eff}$, which quantifies the effective contact area due to curved surfaces.
 We consider the total number of sites as $N_{\rm max} = L^2/s_{\rm hd}$ with the box length $L$, and the average number of receptors per site $\langle n_r \rangle = \pi r_{\rm eff}^2 \rho_r$.
The guest activity is given by $z_g = e^{\beta\mu_{g}} s_{\rm hd} h_0 / \Lambda^3$, with the adsorption layer thickness $h_0=1$. 
The adsorption probability in the simulation is calculated as \(\theta = N_g/N_{\rm max}\), where \(N_g\) is the average number of guest particles adsorbed.

To quantify the selectivity of nanoparticles adsorption with respect to the variation of environmental parameter $\chi$, we generalize the definition of the multivalent selectivity in Ref. \cite{martinez2011designing} as 
\begin{equation}\label{eq:alpha}
    \alpha_\chi = \frac{{\rm ~d} \log  \theta }{ {\rm ~d} \log {\chi}}.
\end{equation} 
When \( \alpha_{\chi} > 1 \), the adsorption increases superlinearly with the environmental parameter \( \theta \sim \chi^{\alpha_{\chi}} \). Similarly, a negative response \( \alpha_{\chi} < -1 \)  signifies that nanoparticles desorb superlinearly. Both cases are defined as $\chi$-dependent superselectivity.

\begin{figure}[!t]
\centering
     \includegraphics[width=0.48\textwidth]{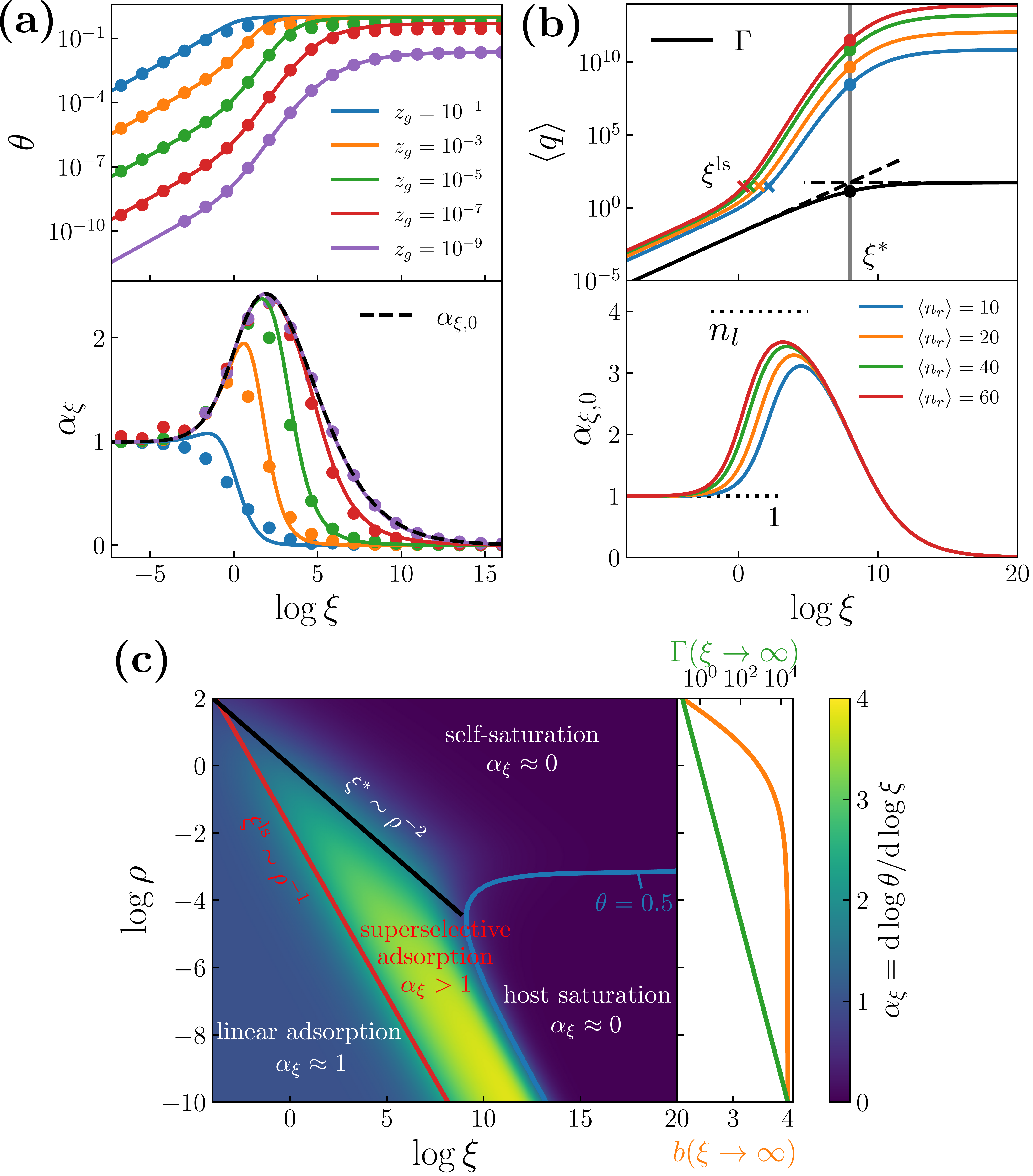}
    \caption{\label{fig2} \textbf{$\xi$-dependent superselectivity.} 
    (a)	$\theta$, $\alpha_\xi$ and $\alpha_{\xi,0}$ as functions of $\log \xi$ at different $z_g$. 
    	Symbols are simulation results with curves the theoretical prediction. 
    	Here $\rho = e^{-2}$, $\rho_r=12.73$ and $r_{\rm eff}=0.5$.
    (b) Theoretical prediction of $\langle q \rangle$, $\Gamma$ (upper black solid line) and $\alpha_{\xi,0}$ as functions of $\log \xi$ at $\rho = e^{-4}$ and various $\langle n_r \rangle$. 
    	Dots locate the self-saturation point $\xi^\ast$, which is the crosspoint between the dashed black lines $\Gamma = {\xi\xi_{\rm cnf}\rho}$ (lower $\xi$) and $\Gamma =  {\xi_{\rm cnf} }\rho^{-1} $ (higher $\xi$). 
    	$\times$ indicate the crossover points $\xi^{\rm ls}$ from linear to superlinear response.
    (c) Left: $\xi$-dependent selectivity $\alpha_\xi$ as a function of $\log \xi$ at different $\log \rho$ and $z_g = 10^{-9}$. Red, black and blue curves indicate the location of $\xi^{\rm ls}$, $\xi^\ast$ and $\theta=0.5$, respectively.
    	Right: at strong bridging limit, 
    	$b(\xi\to\infty)$ (yellow) and $\Gamma(\xi\to\infty)$ (green) as functions of $\log \rho$.
		In all figures, unless otherwise specified, $\langle n_r \rangle=10$, $n_l=4$ and $\Delta S_{\rm cnf}=0$. 
		}
\end{figure}

\paragraph*{Bridging strength dependent superselectivity.}
To investigate the superselective response to temperature or individual bridging energy (e.g., different base pair matching~\cite{taton2000scanometric}) of linker-mediated adsorption, we define the environmental parameter as the bridging strength of a single bridge, $\xi := \xi_{l}\xi_{r}$. We focus on the situation where the receptors significantly outnumber the ligands \( \langle n_r \rangle \gg n_l \).
As shown in Fig.~\ref{fig2}a, with increasing $\xi$, $\theta$ increases and reaches a plateau after a $\xi$ threshold. Similar to the adsorption in direct binding, the guest nanoparticle activity $z_g$ controls the location of the threshold~\cite{martinez2011designing}. When $z_g$ is large enough, e.g., $z_g = 10^{-1}$, only individual bridging can saturate the host, and no $\xi$-dependent superselectivity exists. 
We note that the host saturation is due to not enough available sites on the host substrate, i.e., $\theta > 0.5$.
At a constant adsorption $\theta$, compensating for a lower \(z_g\) with stronger effective interactions, namely an increased number of bridges, results in the gradual emergence of superselectivity. This process is accompanied by an increased peak in \(\alpha_{\xi}\) (Fig~\ref{fig2}a).
Finally, \(\alpha_\xi\) converges to the zero-activity selectivity \(\alpha_{\xi,0}={{\rm d} \log \langle q \rangle}/{{\rm d} \log \xi}\), where \(\langle q \rangle\) is the adsorbed state partition function under the Poisson distribution of receptors.

Intriguingly, when $z_g$ is small, e.g., $z_g=10^{-9}$,  $\theta(\xi \rightarrow \infty)$ is significantly smaller than 1 (Fig.~\ref{fig2}a), i.e., the host cannot be fully occupied in the limit of strong bridging or zero temperature, which is different from the direct binding of multivalent nanoparticles. 
To understand this, we plot $\langle q\rangle$ and  $\alpha_{\xi,0}$ as functions of $\xi$ in Fig.~\ref{fig2}b.
We find that the $\xi$-response can be divided into three parts. 
At the weak bridging $\xi < \xi^{\rm ls}$, the bridging strength is too weak to induce cooperative effects, and the linear response $\alpha_{\xi,0}\approx 1$ appears. 
When $\xi > \xi^{\rm ls}$, stronger bridging strength triggers the formation of multiple bridges, a power-law response $\alpha_{\xi,0}\approx n_l$ appears, indicating the $\xi$-dependent superselectivity. Here, larger multivalency naturally promotes the number of bridges and enhances the selectivity. 
The bridging strength \rt{$\xi^{\rm ls}= 1/(\tilde{n} \xi_{\rm cnf}\rho)$} indicates the crossover from linear to superlinear response, \rt{where $\tilde{n}=\langle n_r \rangle/n_l^{1/(n_l-1)}$ signifies the contribution from the multivalency of both guests and hosts, and increases with increasing $n_{r/l}$} \rt{(see SM Sec.~II~\cite{supinfo2})}.
Moreover, when $\xi$ further increases, $\langle q \rangle$ saturates with no selectivity $\alpha_{\xi,0}\approx 0$. At the large $\xi$ limit, the relative bridging strength $\Gamma(\xi \to \infty) \approx {\xi_{\rm cnf} }\rho^{-1} \sim \xi^{0}$ is dictated solely by entropy~\cite{zilman2003entropic,xia2020linker}. 
The reason is that all receptors and ligands are bound by linkers at this limit. Consequently, the energy remains constant, 
and the bridge formation is a result of competition between entropy lost of forming a bridge and the entropy gain of releasing a free linker~\cite{xia2020linker}.
 As shown in Fig.~\ref{fig2}b, the superlinear $\xi$-response terminates at $\xi^\ast \sim \rho^{-2}$ \rt{(see SM Sec.~II~\cite{supinfo2})}.
This entropy induced \emph{self-saturation} effect makes $ \theta $ unable to reach 1 at the large bridging strength limit, which qualitatively differs from the host saturation ($\theta > 0.5$).

Experimentally, the detectable range of \( \theta \) is limited by measurement techniques or environmental noises. For example, in scanometric DNA array detections, particle probes must polymerize to form network structures, inducing noticeable color shifts~\cite{taton2000scanometric}. Lower \( z_g \) makes the system unable to reach the detectable limit, eliminating the detectable superselectivity. Thus, superselectivity can be only observed at certain range of $z_g$, different from the direct binding where low \( z_g \) typically enhances the superselectivity~\cite{wang2012selectivity,phan2023bimodal}.

Next, we investigate the condition of $\xi$-dependent superselectivity appearing at different $\rho$.
The contourplot of $\alpha_{\xi}$ in $\log \xi$-$\log \rho$ plane at $z_g = 10^{-9}$ is shown in Fig.~\ref{fig2}c, where superselectivity ($\alpha_{\xi} > 1$) emerges within the regions between $\xi^{\rm ls} \sim \rho^{-1}$, $\xi^\ast \sim \rho^{-2}$, and under the host saturation $\theta<0.5$.
Interestingly, with decreasing $\rho$, superselectivity increases.
The reason is that, due to the self-saturation effect, the maximum bridge number $b(\xi \to \infty)$ is $\rho$-dependent, instead of the constant $n_l$ in direct binding~\cite{martinez2011designing}. Smaller $\rho$ induces larger $\Gamma(\xi \to \infty)$ to form more bridges. The bridge number, which can be seen as the effective valence, determines the cooperativity triggering the superselectivity. 
%\rt{As indicated by the green and orange lines in Fig.~\ref{fig2}c, both \(\Gamma\) and \(b\) are small at high linker densities, suggesting that the bounds of superselectivity are governed by self-saturation. Conversely, at lower linker densities, there is an increase in \(\Gamma(\xi \to \infty)\) and \(b(\xi \to \infty)\), which implies that the bounds of superselectivity are dictated by the linker entropy.}

\begin{figure*}[!t]
\centering
     \includegraphics[width=1.0\textwidth]{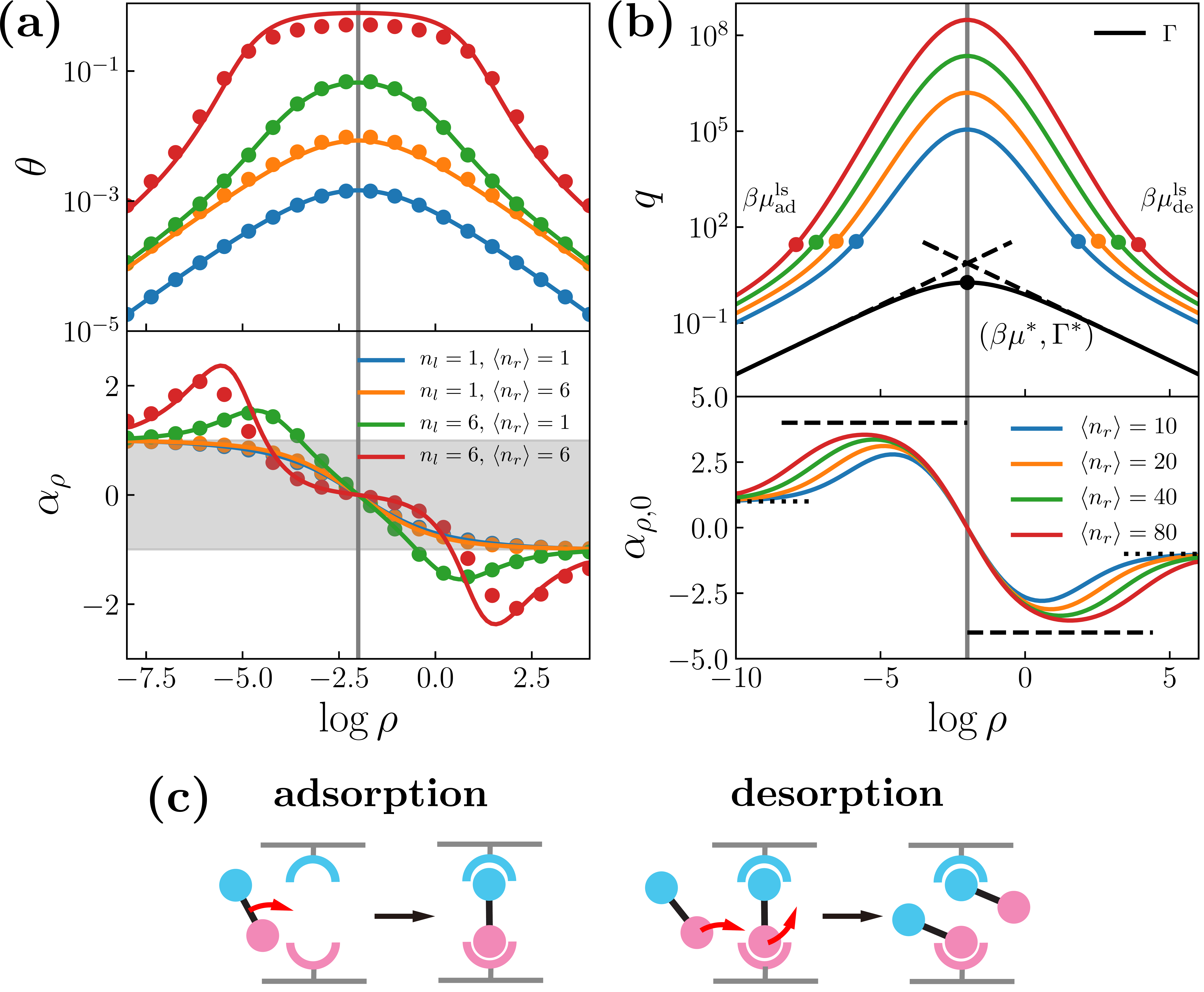}
    \caption{\label{fig3}\textbf{$\rho$-dependent superselectivity.}     
     (a) $\theta$ and $\alpha_{\rho}$ as functions of $\log \rho$ for mono- and multivalent ligands with $\langle n_r \rangle = 1$ and $6$. Grey area indicates no superselectivity. Here $\xi_l=\xi_r= e^{2}$, $z_g = 10^{-3}$ and $r_{\rm eff}=0.5$. 
     (b) Theoretical prediction of $\langle q \rangle$, $\Gamma$ (upper black solid line) and $\alpha_{\rho,0}$ as functions of $\log \rho$ at $\xi_l=\xi_r= e^{-2}$ with different $\langle n_r \rangle$. 
     Black dot locates the adsorption-desorption transition point $\rho^\ast$, the crosspoint between the lines $\Gamma = {\xi\xi_{\rm cnf}\rho}$ (adsorption regime) and $\Gamma =  {\xi_{\rm cnf} }\rho^{-1} $ (desorption regime). Other colored symbols indicates the crossover points $\rho^{\rm ls}_{\rm ad}$ and $\rho^{\rm ls}_{\rm de}$ from linear to superlinear responses in adsorption and desorption regimes, respectively.
     (c) Illustration of adsorption and desorption with increasing linker density where bridges are insufficient (upper) and excessive (lower).
     (d) Left: $\rho$-dependent selectivity $\alpha_\rho$ as a function of $\log \rho$ at different $\log \xi$ and $z_g = 10^{-9}$. Red, black, blue and orange curves indicate the location of $\rho^{\rm ls}_{\rm ad}$, $\rho^\ast$, $\rho^{\rm ls}_{\rm de}$ and $\theta=0.5$, respectively.
    	Right: at $\rho^\ast$, $b(\rho^\ast)$ (green) and $\Gamma(\rho^\ast)$ (purple) as functions of $\log \xi$.
     }
\end{figure*}

\paragraph*{Linker density dependent superselectivity.}
Modifying linker density enables specific object binding in environments with temperature constraints.
As shown in Fig.~\ref{fig3}a, with increasing linker density \( \rho \), $\theta$ initially increases and then decreases in all cases. The peak location in $\theta$ corresponding to the targeted linker density is \( \rho^\ast = \xi^{-1/2}\) \rt{(see SM Sec.~III~\cite{supinfo2})}, and below and above this peak are defined as the adsorption and desorption regimes, respectively. As shown in Fig.~\ref{fig3}b, we find that the adsorption-desorption transition originates from the non-monotonic dependence of $\Gamma$ on $\rho$. When $\rho<\rho^\ast$, we have $\Gamma \approx {\xi \xi_{\rm cnf}\rho}$, where the bridges are insufficient, and the adsorption involves sequential insertions of linkers and formation of bridges (upper panel of Fig.~\ref{fig3}c). Conversely, in the desorption regime, we have $\Gamma \approx {\xi_{\rm cnf} }\rho^{-1}$, where the bridges is excessive, and the desorption process involves free linkers attacking bridges and forming dangling bonds (lower panel of Fig.~\ref{fig3}c).
Various linker-mediated systems exhibit this non-monotonic dependence on linker density, including sandwich immunoassays~\cite{tate2004interferences}, DNA linked colloids~\cite{xia2020linker,lowensohn2019linker}, (bio)-polymeric networks~\cite{lei2020entropy,banerjee2017reentrant,hong2020behavior,ren2022uncovering} and cell-mediated aggregation~\cite{dias2020modeling}, with theoretical explanations~\cite{xia2020linker,douglass2013comprehensive}. 

We consider the linker density $\rho$ as the environmental parameter in Eq.~\ref{eq:alpha} with the corresponding selectivity parameter $\alpha_{\rho}=\rm{d}\log \theta / \rm{d}\log \rho $. 
As shown in Fig.~\ref{fig3}a, the superselective adsorption and desorption $|\alpha_{\rho}| >1$ occur when the guest nanoparticles are multivalent, e.g., $n_l=6$, where multiple bridges can form. Particularly, the superselectivity also occurs at $\langle n_r\rangle=1$, which implies that the spatial inhomogeneity of receptors promotes $\rho$-dependent selectivity, since it involves larger $n_r$ and $q$. 
We use zero-activity selectivity $\alpha_{\rho,0}={{\rm d} \log \langle q\rangle} / {{\rm d} {\log \rho}}$ to analyze the origin of $\rho$-dependent superselectivity. As shown in Fig.~\ref{fig3}b, we obtain four modes of $\langle q \rangle$ dependence  \rt{(see SM Sec.~II and III~\cite{supinfo2})}.
At the limit of both low and high linker densities, weak bridging $\Gamma\to 0$ can not trigger cooperativity, and $|\alpha_{\rho,0}| \approx 1$.
\rt{The strongest relative bridging strength $\Gamma(\rho^\ast)$ increases with $\xi_{l/r}$, thus,} 
large $\Gamma$ triggers multiple bridges to form, and the superselectivity characterized by a power-law dependence $|\alpha_{\mu,0}| \approx n_l$ appears.
The crossover linker densities from linear to superlinear response in the adsorption and desorption regime \rt{are} \rt{$\rho_{\rm ad }^{\rm ls}={  1 /(\tilde{n}} {\xi\xi_{\mathrm{cnf}})}$} and \rt{$\rho_{\rm de}^{\rm ls} = {\tilde{n} \xi_{\mathrm{cnf}} }$}, respectively.
Moreover, the maximum $|\alpha_{\mu,0}|$ increases with increasing the receptor density on substrate $\langle n_r \rangle$ (Fig.~\ref{fig3}b).
Hence, the linker-mediated multivalent adsorption provides a clean and shutterlike ``off-on-off'' response targeting a specific linker density or density window. 

We demonstrate how to design \(\rho\)-dependent superselectivity in a system by adjusting the bridging strength \(\xi\). As shown in Fig.~\ref{fig3}d, the superselective regime is confined by the lower and upper bounds \(\rho_{\rm ad }^{\rm ls} \sim \xi^{-1}\) and \(\rho_{\rm de}^{\rm ls} \sim \xi^0\), with the targeted linker density \(\rho^\ast \sim \xi^{-1/2}\) dividing the regime into adsorption and desorption parts. Counterintuitively, larger bridging strength promotes higher selectivity, opposite to the trend in direct binding, where weaker individual binding typically enhances selectivity~\cite{martinez2011designing,scheepers2020multivalent,linne2021direct}. The reason is that stronger bridging leads to a higher relative bridging strength \(\Gamma(\rho^\ast)\), resulting in more bridges \(b(\rho^\ast)\) formed, as shown by the purple and green curves in Fig.~\ref{fig3}d. Furthermore, due to the self-saturation effect, the adsorption is even tunable at the large \(\xi\) (zero temperature) limit.  Increasing \(\rho\) induces superselective desorption at the limit, driven solely by entropy.

Combining both $\xi$- and $\rho$-dependent superselectivity, we propose a general design rule for the superselective adsorption of linker-mediated multivalent nanoparticles. When bridging is insufficient, i.e., $\xi \rho^{2}<1$, both $\xi$- and $\rho$-superselectivity appears at \rt{$\xi \rho > 1  / (\tilde{n} \xi_{\rm cnf})$}. In the excess bridging region, i.e., $\xi \rho^{2}>1$, entropy dictates the adsorption and the corresponding superselectivity. The entropy induced self-saturation effect eliminates the $\xi$ dependence, and the superselective desorption appears with increasing $\rho$ when \rt{$\rho < \tilde{n} \xi_{\rm cnf}$}. We note that the guest nanoparticle activity $z_g$ 
can shift the location of saturation threshold, which can be experimentally controlled by adjusting the nanoparticle density and/or introducing extra repulsion or attraction between nanoparticles and the substrate~\cite{phan2023bimodal}. 
Moreover, when the bridging is asymmetric, i.e., $\Delta \xi = \xi_l / \xi_r \neq 1$, peaks of both $\xi$- and $\rho$-dependent selectivity decrease \rt{(see SM Sec.~II~\cite{supinfo2})}. The asymmetric bridging introduces a plateau with indiscriminate adsorption of $\rho$, of which the width is $\Delta \xi$ (Fig.~S1\rt{~\cite{supinfo2}}).
This explains that the melting temperature of crystals consisting of asymmetric DNA linked colloids remains constant in certain range of linker density~\cite{lowensohn2020self}.

\paragraph*{Discussion.} 
The interplay between energy and entropy in linker-mediated interactions results in rich phase behavior~\cite{xia2020linker}, making the superselectivity arising from multivalency intricate and distinct from the direct binding. 
Here, we have investigated both the bridging strength and linker density dependent superselectivity in linker-mediated multivalent nanoparticle adsorption. We theoretically estimate the bridge number and adsorption probability and elucidate the origin of superselectivity in various scenarios. We find that strong relative bridging strength \( \Gamma \) and multivalency lead to multiple bridges formed between host and guest nanoparticles, yielding  superselectivity. We estimate the parameter bounds as guiding principles for designing the linker-mediated superselectivity.
Our theory suggests that for any experimental parameter $\chi$, if $\Gamma \sim \chi$, then $\chi$-dependent superselectivity occurs.
Apart from applications in biosensing, drug delivery and self-assembly, our finding may also shed lights on the understanding of fundamental physics in biological systems, such as client recruitment in biomolecular condensates~\cite{banani2016compositional,christy2021control,yang2020g3bp1}.

\begin{acknowledgments}
{The authors thank Tianran Zhang for helpful discussions.} This work is supported by the Academic Research Fund from the Singapore Ministry of Education (RG59/21 and MOE2019-T2-2-010), and the National Research Foundation, Singapore, under its 29th Competitive Research Program (CRP) Call (Award No. NRF-CRP29-2022-0002).
\end{acknowledgments}

\bibliography{abbr}% apssamp make_abbr/abbr

\end{document}